\documentstyle[12pt,a4]{article}
\newlength{\extraspace}
\newlength{\extraspaces}
\newcommand{\be}{\begin{equation}
\addtolength{\abovedisplayskip}{\extraspaces}
\addtolength{\belowdisplayskip}{\extraspaces}
\addtolength{\abovedisplayshortskip}{\extraspace}
\addtolength{\belowdisplayshortskip}{\extraspace}}
\newcommand{\ee}{\end{equation}}

\newcommand{\ba}{\begin{eqnarray}
\addtolength{\abovedisplayskip}{\extraspaces}
\addtolength{\belowdisplayskip}{\extraspaces}
\addtolength{\abovedisplayshortskip}{\extraspace}
\addtolength{\belowdisplayshortskip}{\extraspace}}
\newcommand{\ea}{\end{eqnarray}}

\newcommand{\nonu}{\nonumber \\[.5mm]}
\newcommand{\A}{&\!\!\!}

\def\thesection {\S {\arabic{section}}}
\newcommand{\newsection}[1]{
\vspace{7mm}
\pagebreak[3]
\addtocounter{section}{1}
\setcounter{subsection}{0}
\setcounter{footnote}{0}
\begin{center}
{\large {\bf \thesection. #1}}
\end{center}
\nopagebreak
\medskip
\nopagebreak
\hspace{3mm}}

\begin{document}


\thispagestyle{empty}

\vspace{.6cm}

\begin{large}

\centerline{\bf Equivalence Principle in the New General Relativity}

\end{large}

\hspace{2cm}


Takeshi SHIRAFUJI$\,{}^\ast$, Gamal G.L. NASHED$\,{}^\ast$
$\!$\footnote[2]{\,Permanent address: Mathematics Department, 
Faculty of Science, Ain Shams University, Cairo, Egypt.} 
and Yoshimitsu KOBAYASHI$\,{}^\ast$

\centerline{${}^\ast$\ {\it Physics Department, Faculty of Science, 
Saitama University, Urawa 338}}

\bigskip


\thispagestyle{empty}

\hspace{2cm}
\\
\\
\\
\\
\\
\\
\\
\\

We study the problem of whether the active gravitational mass of an
isolated system is equal to the total energy in the tetrad theory of 
gravitation. The superpotential is derived using the gravitational 
Lagrangian which is invariant under parity operation, and applied to an
exact spherically symmetric solution. Its associated energy is found 
equal to the gravitational mass.  The field equation in vacuum is also solved at far distances under the assumption of spherical symmetry. Using the most 
general expression for parallel vector 
fields with spherical symmetry, we find that the equality between the  
gravitational mass and the energy is always true if the parameters of the 
theory $a_1$, $a_2$ and $a_3$ satisfy the
 condition, $(a_1+ a_2) (a_1-4a_3/9)\neq0$. In the two special cases where 
either $(a_1+a_2)$ or 
$(a_1-4a_3/9)$ is vanishing, however, this equality is not satisfied for the 
solutions when some components of the parallel vector fields tend to zero 
as $1/\sqrt{r}$ for large $r$.


\newpage
\newsection{Introduction}
It was shown by ${\rm M\phi ller}^{1)}$ 
that a tetrad description of gravitational field allows a more satisfactory
treatment of the energy-momentum complex than general relativity (GR).
The 
Lagrangian formulation of the theory was given by Pellegrini and 
${\rm Plebanski.}^{2)}$
 Hayashi and ${\rm Nakano}^{3)}$ 
independently formulated the tetrad
theory of gravitation as a gauge theory of the spacetime translation group.
In the earlier attempts, admissible Lagrangians were limited by the 
assumption that the field equation has the 
Schwarzschild solution. ${\rm M\phi ller}^{4)}$ later suggested to abandon 
this assumption  and to look for a wider class of Lagrangians.
 ${\rm Meyer}^{5)}$ formulated the tetrad theory as a special case of 
Poincar$\acute{e}$ gauge ${\rm theory.}^{6),7)}$ ${\rm S\acute{a} ez}^{8)}$ 
generalized the theory into a scalar-tetrad theory. 

Hayashi and ${\rm Shirafuji}^{9)}$ studied the geometrical and observational 
basis of the tetrad  theory of gravitation. Geometrically the
tetrad fields are identified with the parallel vector fields defined by the
underlying absolute parallelism. Incidentally they 
gave the name, new general relativity (NGR), to the theory of gravitation based on absolute parallelism, since ${\rm Einstein}^{10)}$ was the first to 
introduce the notion of absolute parallelism in physics.
 They assumed the Lagrangian 
 with three unknown parameters, denoted by $a_1$, $a_2$ and $a_3$.
 In order to reproduce the correct Newtonian limit, the first two parameters 
should satisfy a condition called the Newtonian
approximation condition, which allows us to express these two parameters in 
terms of an unknown 
dimensionless parameter $\epsilon$.
An exact, vacuum solution of the gravitational field equation was found for 
${\it static}$, spherically symmetric case, where the parallel vector fields 
take a diagonal form. This solution describes the spacetime around a mass point  located at the origin, and will be referred to as the "exact spherically 
symmetric solution" hereafter. Then comparison with solar-system
 experiments showed that the value of $\epsilon$ should be very small. By contrast only an upper bound has  been estimated for the remaining parameter 
${\it a}_3.^{9),11)}$  The singularity problem of the exact 
solution has been ${\rm studied.}^{12),13)}$
If $(a_1+a_2)$=0, then the theory reduces to the one studied by Hayashi and
${\rm Nakano}^{3)}$ and ${\rm M\phi ller.}^{4)}$

The equation of motion for a test particle  was discussed  based  on the 
response equation of the energy-momentum tensor of matter ${\rm fields.}^{9)}$
In particular, when the intrinsic spin of the fundamental particles 
constituting the test 
particle can be ignored, the response equation reduces to the covariant
conservation law of GR, and hence the world line of the test particle is a
 geodesics of the metric defined by the parallel vector fields. Accordingly, 
for spinless test particles the passive gravitational mass is equal to the 
inertial mass. 

In GR the equality between passive gravitational mass and inertial mass is 
predicted to be valid also for massive bodies like planets which contains 
appreciable fraction (about $5\times10^{-10}$ for the Earth) of gravitational 
self-energy. This equality means that gravity pulls on gravitational binding 
energy 
of a massive body just as it does on other forms of mass-energy, and is referred to as the strong equivalence ${\rm principle.}^{14)}$ ${\rm Nordtvedt}^{15)}$ 
developed the parametrized post-Newtonian (PPN) formalism and calculated the 
diviation from the unity of the ratio of gravitational mass to inertial mass 
in terms of the PPN parameters. Analysis of the lunar laser ranging (LLR) over 
past 24 years has confirmed that the Earth and the Moon accelerate equally to 
the sun within fractional difference less than $5\times10^{-13}$, which yields the
Nordtvedt parameter $\eta=-0.0010 \pm 0.0010$.$^{16),17),18)}$ This parameter 
$\eta$ ($\equiv 0$ in GR) can be expressed as  $\eta=4\beta-\gamma-3$, 
in fully conservative theory which possesses a full complement of the 
post-Newtonian conservation laws: energy, momentum, angular momentum and 
center-of-mass motion.$^{14)}$ Here $\beta$ and $\gamma$ are the so-called 
Eddington-Robertson parameters.$^{19),20)}$  

Thus, if NGR is a fully conservative theory in the above sense, then the
Nordtvedt parameter $\eta$ vanishes, since the Eddington-Robertson parameters
 are expressed as $\beta=1-\epsilon/2$ and 
$\gamma=1-2\epsilon$  in terms of the dimensionless parameter $\epsilon$ 
mentioned above. It can be shown that energy, momentum and angular momentum  
are conserved in NGR. We must be careful not to conclude that $\eta=0$, however, since NGR is not a metric theory: Accordingly we are now trying to develop its PPN formalism.

It is the purpose of this paper to study a different aspect of the equivalence 
principle, namely  the problem of whether or not the active gravitational 
mass (or simply the gravitational mass) of an isolated system is equal
 to its inertial mass, i.e. the total energy divided by the square of the 
velocity of light.\footnote{Since we use the unit $c=1$, we need not draw a 
distinction between the inertial mass and the total energy}  As is well known,
 this problem is settled affirmatively in 
${\rm GR.}^{21)}$ In the case of $(a_1+a_2)=0$ Mikhail et
${\rm al.}^{22)}$ calculated the energy of two spherically symmetric
solutions, and found that the energy in one of the two solutions does not
coincide with the gravitational mass.
Shirafuji et ${\rm al.}^{23)}$ extended the calculation to all the 
 stationary asymptotically flat solutions with spherical symmetry,
 dividing them into two classes:
The one in which the components, $({b^a}_{0})$ and $({b^{(0)}}_{\alpha})$, 
of the parallel vector fields 
$({b^k}_{\mu})$  tend to zero faster than $1/\sqrt{r}$ for large $r$ and the 
other in which those components go to zero like
 $1/\sqrt{r}$ .\footnote{In this paper Latin indices $(i,j,k,...)$ represent the vector number, which runs from $(0)$ to $(3)$, while Greek indices $(\mu,\nu,\rho, ...)$ represent the world-vector components running from 0 to 3. The 
spatial part of Latin indices are denoted by $(a,b,c,...)$, while that of Greek indices by $(\alpha, \beta,\gamma,...)$.}  It was found that the equality of 
the 
gravitational and inertial masses holds true only in the first class.

In this paper we study the problem for stationary, spherically symmetric systems without making any assumption on the parameters, $a_1$, $a_2$ and $a_3$ other
 than the Newtonian approximation condition.

 We organize this paper as follows. In {\S 2} we give a brief review of the
 NGR. In \S 3 we derive the 
 superpotential of the total energy-momentum complex, and  apply it to the 
exact spherically symmetric solution. In \S 4 we study
stationary spherically symmetric solutions at far distances in the
linear approximation.  
In \S 5 we calculate the energy in the special case of $(a_1-4a_3/9)=0$, 
taking all the leading terms into account beyond the linear approximation. 
The final section is devoted to  conclusion and discussion.

\newsection{A brief review of the NGR}
We assume spacetime to admit absolute parallelism, i.e. to have
a quadruplet of linearly independent parallel vector fields $({b^k}_{\mu})$ 
satisfying
\be
{D_\nu} {b^k}_\mu = {b^k}_{\mu,\nu}-
{\Gamma^\lambda}_{\mu \nu}{b^k}_\lambda=0
\ee
with ${b^k}_{\mu,\nu}= {\partial}_\nu {b^k}_{\mu}$. Solving (1), we get
the nonsymmetric connection
\be
{\Gamma^\lambda}_{\mu \nu} ={b_k}^\lambda {b^k}_{\mu,\nu},
\ee
which defines the torsion tensor as
\be
{T^\lambda}_{\mu \nu} = {\Gamma^\lambda}_{\mu \nu}-{\Gamma^\lambda}_{\nu \mu}
={b_k}^\lambda ({b^k}_{\mu,\nu}-{b^k}_{\nu,\mu}).
\ee
The curvature tensor defined by ${\Gamma^\lambda}_{\mu \nu}$ is identically 
vanishing, however. The metric tensor is given by the parallel vector fields 
as
\be
g_{\mu \nu}= b_{k \mu} {b^k}_ \nu,
\ee
where we raise or lower Latin indices by the Minkowski metric 
$\eta_{i j}$=$\eta^{i j}$ =diag(-1,+1,+1,+1).

Assuming the invariance under \\
a) the group of general coordinate transformations,\\
b) the group of global Lorentz transformations, and \\
c) the parity operation,\\
 we write the gravitational Lagrangian density in the form \footnote{
Throughout this paper we use the relativistic units, $c=G=1$. The Einstein constant ${\kappa}$ is then equal to $8 \pi $. We will denote the symmetric
part by the parenthesis (\ ) and the  antisymmetric part by the square bracket 
[\ ].} 
\be
{\cal L}_G= {\sqrt{-g}\over \kappa} \left[a_1(t^{\mu \nu \lambda}
t_{\mu \nu \lambda})+a_2(v^{\mu} v_{\mu})+a_3(a^{\mu}a_{\mu}) \right],
\ee
where $a_1$, $a_2$ and $a_3$ are 
dimensionless parameters of the theory, \footnote{ The dimensionless parameters $\kappa a_i$ of Ref.9) are here denoted by $a_i$ for convenience.} and
\ba
t_{\mu \nu \lambda} \A = \A {1 \over 2} 
\left(T_{\mu \nu \lambda}
+T_{\nu \mu \lambda} \right)
+{1 \over 6} \left(g_{\lambda \mu} v_\nu
+g_{\lambda \nu} v_{\mu} \right)-{1 \over 3} g_{\mu \nu} v_\lambda, \\
v_{\mu} \A = \A {T^\lambda}_{\lambda \mu}, \\
a_{\mu} \A = \A {1 \over 6}{\epsilon}_{\mu \nu \rho \sigma} T^{\nu \rho \sigma}
\ea
with ${\epsilon}_{\mu \nu \rho \sigma}$ being the completely antisymmetric 
tensor normalized as ${\epsilon}_{0123}=\sqrt{-g}$.
By applying variational principle to the above Lagrangian, we get the field 
equation:
\be
I^{\mu \nu}= {\kappa}T^{\mu \nu}
\ee
with
\be
I^{\mu \nu}=2{\kappa}[{D}_\lambda F^{\mu \nu \lambda}+ 
v_\lambda F^{\mu \nu \lambda}+H^{\mu \nu}
-{1 \over 2} g^{\mu \nu}L_G],
\ee
where
\ba
F^{\mu \nu \lambda} \A = \A {1 \over 2} b^{k \mu} {\partial L_G \over \partial 
{b^k}_{\nu,\lambda}} \nonu
 \A = \A {1 \over \kappa} \left[ a_1 \left(t^{\mu \nu \lambda}
-t^{\mu \lambda \nu} \right)+a_2 \left(g^{\mu \nu} v^\lambda
-g^{\mu \lambda} v^\nu \right)
-{a_3 \over 3} \epsilon^{\mu  \nu \lambda \rho} a_\rho \right]
=-F^{\mu \lambda \nu},\\
H^{\mu \nu} \A = \A T^{\rho \sigma \mu} 
 {F_{\rho \sigma}}^\nu - {1 \over 2} T^{\nu \rho \sigma} 
{F^\mu}_{\rho \sigma}=H^{\nu \mu},\\
{L_G} \A = \A {{\cal L}_G \over \sqrt{-g}},\\
T^{\mu \nu} \A = \A {1 \over \sqrt{-g}} {\delta {\cal L}_M \over 
\delta {b^k}_\nu} b^{k \mu}.
\ea
Here ${\cal L}_M$ denotes the Lagrangian density of material fields and
$T^{\mu \nu}$ is the material energy-momentum tensor which is 
nonsymmetric in general.
In order to reproduce the correct Newtonian limit, we require the parameters
$a_1$ and $a_2$ to satisfy the condition
\be
a_1+4a_2+9a_1a_2=0
\ee
called the Newtonian approximation ${\rm condition,}^{9)}$  which can be 
solved to give
\be
a_1=-{1 \over 3(1-\epsilon)}, \quad 
a_2={1 \over 3(1-4\epsilon)}
\ee
with $\epsilon$ being a dimensionless parameter. The comparison with 
solar-system experiments shows that $\epsilon$ should be given by$^{\rm 9)}$
\be
\epsilon=-0.004 \pm0.004,
\ee
which we assume throughout this paper.

It is well known that the conservation law in GR is given by
\be
{{T_{GR}}^{\mu \nu}}_{;\nu}= 0,
\ee
where ${T_{GR}}^{\mu \nu}$ is the symmetric material energy-momentum  tensor
of GR and the semicolon denotes covariant derivative with respect to the 
Christoffel symbol. This law does not follow from (9), however.
Instead, we can derive the response equation 
\be
{T^{\mu \nu}}_{;\nu}=K^{\nu \lambda \mu}T_{[\nu \lambda]},
\ee
where $K^{\nu \lambda \mu}$ is the contortion tensor given by
\be
K^{\nu \lambda \mu}={1 \over 2} \left( T^{\nu \lambda \mu}-T^{\lambda \nu \mu}
-T^{\mu \nu \lambda} \right)=-K^{\lambda \nu \mu}.
\ee
The antisymmetric part $T^{[\mu \nu]}$ is due to the contribution from the
intrinsic spin of fundamental spin-$1/2$ particles. For macroscopic test
particles  for which the effects due to intrinsic spin 
can be ignored, their energy-momentum tensor can be supposed to be
symmetric and satisfy (18). The equation of motion for such
macroscopic test particles is then the geodesic equation of the metric.

In ${\it static}$, spherically symmetric spacetime the parallel vector fields 
take a diagonal form, and the field equation (9) 
can be exactly solved in vacuum to give
\be
\left( {b^k}_\mu \right) = 
\left(
\matrix{
{\displaystyle \left( 1-\displaystyle{m \over pr} \right)^{p/2}  \over
\left( 1+\displaystyle{m \over qr} \right)^{q/2}} & 
0 \vspace{3mm} \cr 
0 & 
\displaystyle \left( 1-\displaystyle{m \over pr} \right)^{(2-p)/2}
\displaystyle \left( 1+\displaystyle{m \over qr} \right)^{(2+q)/2} 
{\delta^a}_\alpha \cr}\right),
\ee
where $m$ is a constant of integration, and  the constants $p$ and $q$ are given by
\be
p={2 \over (1-5\epsilon)}[(1-5\epsilon+4\epsilon^2)^{1 \over 2}-2\epsilon], 
\quad q={2 \over (1-5\epsilon)}[(1-5\epsilon+4\epsilon^2)^{1 \over 2}+2\epsilon].
\ee
In spherical polar coordinates (21) gives the line-element 
\be
ds^2=-{\displaystyle \left( 1-{m \over pr} \right)^p \over 
\displaystyle \left( 1+{m \over qr} \right)^q} 
dt^2+\left( 1-{m \over pr} \right)^{2-p} \left( 1+{m \over qr} \right)^{2+q} 
\left [dr^2+r^2(d\theta^2 +{\rm sin^2} \theta d\phi^2) \right].
\ee
>From the asymptotic behavior of the component $g_{0 0}$ of the metric tensor,
 the constant ${\it m}$ can be identified with the 
gravitational mass of the central gravitating system. It is clear that if
the parameter $\epsilon$ vanishes, the line-element (23) coincides with the 
Schwarzschild solution written in the isotropic coordinates.

\newsection{Superpotential of the NGR and calculation of the energy}
The following identity can be derived from the invariance of ${\cal L}_G$ 
under general coordinate ${\rm transformations:}^{1),24)}$
\be
 -{\delta {\cal L}_G \over 
\delta{{b^k}_{\nu}}}{{b^k}_{\mu}} 
-{\partial {\cal L}_G \over 
\partial{{b^k}_{\lambda,\nu}}}{{b^k}_{\lambda,\mu}} 
+{\delta^\nu}_\mu{\cal L}_G-{\partial_{\lambda}} \left({\partial {\cal L}_G 
\over \partial{{b^k}_{\nu,\lambda}}}{{b^k}_{\mu}} \right)  \equiv 0.
\ee
When $({b^k}_{\mu})$ satisfies the gravitational field equation (9), this 
implies 
\be
\sqrt{-g} \left( {T_\mu}^\nu+{t_\mu}^\nu \right)=\partial_{\lambda} 
\left( 2\sqrt{-g} {F_\mu}^{\nu \lambda} \right),
\ee
where ${t_\mu}^\nu$ is the canonical energy-momentum complex of the 
gravitational field
\be
\sqrt{-g}{t_\mu}^\nu =
-{\partial {\cal L}_G \over \partial{{b^k}_{\lambda,\nu}}}
{{b^k}_{\lambda,\mu}}+{\delta_\mu}^\nu {\cal L}_G .
\ee
The total energy-momentum complex is then defined by 
\be
{\cal M_\mu}^\nu = \sqrt{-g}({T_\mu}^\nu+{t_\mu}^\nu)= 
{{\cal U_\mu}^{\nu \lambda},_{\lambda}}
\ee
with ${\cal U_\mu}^{\nu \lambda}$ being the superpotential  
\footnote{The Lagrangian used by ${\rm M\phi ller}^{4)}$ is different from that  used by Hayashi and ${\rm Shirafuji}^{9)}$ by a factor (-2). Accordingly, the  definition (28) is different from that of M$\phi$ller by a factor (-2).} 
\be
{\cal U_\mu}^{\nu \lambda} = 2 \sqrt{-g}{F_\mu}^{\nu \lambda}.
\ee
Since the tensor ${F_\mu}^{\nu \lambda}$ is antisymmetric with respect to
$\nu$ and $\lambda$, the ${\cal M_\mu}^\nu$ of (27) satisfies ordinary 
conservation law, 
\be
\partial_{\nu}{\cal M_\mu}^\nu=0.
\ee
The total energy is now given by
\be
E=-\int {{\cal M}_0}^0 d^3 x=
- \lim_{r \rightarrow \infty}\int_{r=constant} 
{{\cal U}_0}^{0 \alpha} n_\alpha dS, 
\ee
where $n_\alpha$ is the outward unit 3-vector normal to the surface element 
$dS$. 

Let us calculate the superpotential by writing the Lagrangian (5) in the 
form
\be
{\cal L}_G={\sqrt{-g} \over \kappa} \left[ a_1 L^{(1)}+a_2L^{(2)}+a_3 L^{(3)} 
\right],
\ee
where  $L^{(1)}= t^{\lambda \mu \nu}t_{\lambda \mu \nu}$, 
$L^{(2)}= v^\mu v_\mu$ and $L^{(3)}= a^\mu a_\mu$. Writing (11) in the form
\be
{F_\mu}^{\nu \lambda}={1 \over \kappa} \left[ a_1 {{F^{(1)}}_\mu}^{\nu \lambda} +a_2 {{F^{(2)}}_\mu}^{\nu \lambda}+a_3 {{F^{(3)}}_\mu}^{\nu \lambda} \right],
\ee
we get
\ba
{{F^{(1)}}_\mu}^{\nu \lambda} \A = \A 
{1 \over 2} \left[ 2{T_\mu}^{\nu \lambda}+{T^{\lambda \nu}}_{\mu}-
{T^{\nu \lambda}}_{\mu}-({\delta_\mu}^{\nu}v^\lambda-
{\delta_\mu}^{\lambda}v^\nu) \right], \\
{{F^{(2)}}_\mu}^{\nu \lambda} \A = \A
 \left( {\delta_\mu}^{\nu}v^\lambda-{\delta_\mu}^{\lambda}v^\nu \right), \\
{{F^{(3)}}_\mu}^{\nu \lambda} \A = \A
-{1 \over 9}
\left[ {T_\mu}^{\nu \lambda}-{T^{\lambda \nu}}_{\mu}+
{T^{\nu \lambda}}_{\mu}\right],
\ea
where ${{F^{(1)}}_\mu}^{\nu \lambda}$, 
${{F^{(2)}}_\mu}^{\nu \lambda}$ and 
${{F^{(3)}}_\mu}^{\nu \lambda}$
correspond to $L^{(1)}$, $L^{(2)}$ and $L^{(3)}$, respectively.
So with the help of (28) the superpotential of the NGR can be written as
\ba
{{\cal U}_\mu}^{\nu \lambda} \A = \A {2 \sqrt{-g} \over \kappa} \Biggl[ 
\left( a_1-{a_3 \over 9} \right) {T_\mu}^{\nu \lambda}
+\left( {a_1 \over 2}+{a_3 \over 9} \right)
 \left( {T^{\lambda \nu}}_{\mu}-{T^{\nu \lambda}}_\mu \right) \nonu
\A  \A  \qquad \qquad \qquad
-\left( {a_1 \over 2} - a_2 \right)
\left( {\delta_\mu}^\nu v^\lambda-{\delta_\mu}^\lambda v^\nu \right) \Biggr].
\ea

As an example, let us apply (36) to the exact solution (21).
The appropriate components ${{\cal U}_0}^{0 \alpha}$ are given by

\ba
{{\cal U}_0}^{0 \alpha} \A= \A-{m \over 2 \kappa pq}{n^\alpha \over r^2} 
\Biggl[ (p-q-8)(a_1-2a_2){m \over r}+4(p-q)
( a_1-2 a_2) 
 \nonu 
\A \A \qquad \qquad \qquad
+4pq(2 a_1-a_2)+3(p-q){a_1 m \over r} \Biggr]
\ea
with $p$ and $q$ given by (22). Using (37) in (30), we get

\ba
E \A= \A {8\pi m \over \kappa pq} \left[ (p-q)(a_1 -2 a_2)+pq(2 a_1 -a_2) 
\right] \nonu%
\A = \A  m \left[ (2 a_1 - a_2)-2\epsilon(a_1 -2 a_2) \right]=m,
\ea
where the relation $(p-q)/pq=-2\epsilon$, which follows from (22),
 is used in the second equality and (16) is employed in the last one.
This means that the total energy of the exact solution is just the same as the
 gravitational mass of the central gravitating system.

\newsection{The spherically symmetric parallel vector fields and its energy
 in the linear approximation} 
The solution (21) is the only exact spherically symmetric solution in vacuum 
that we know at present, and it is not clear to us whether there exist any other spherically symmetric solutions in vacuum. In view of this let us consider a 
wider class of spherically symmetric solutions at far distances from the source.
Consider an isolated, gravitating system with spherical symmetry, and restrict 
attention to the weak field far from the source. The most general 
form of the parallel vector fields is given in the Cartesian coordinates 
${\rm by}^{25)}$
\be
\left({b^k}_\mu \right)= 
\left(
\matrix{
C(r) & G(r) n^\alpha \vspace{3mm} \cr 
H(r) n^a & {\delta^a}_\alpha D(r) + E(r) n^a n^\alpha 
+ F(r) \epsilon_{a \alpha \beta} n^\beta \cr
}
\right),
\ee
where the two real functions ${\it C(r)}$ and  ${\it D(r)}$
 are supposed to approach 1
at infinity, while the remaining four real functions, ${\it E(r)}$, 
${\it F(r)}$, ${\it G(r)}$ and ${\it H(r)}$, must tend to zero there. 
 Here we define the radial unit vector $n^a$ and $n^\alpha$ by
\be
n^\alpha={x^\alpha \over r}={\delta^\alpha}_a n^a,
\ee
without making distinction between upper and lower indices.
 Using the freedom to redefine the radial
coordinate ${\it r}$, we can eliminate the function ${\it E(r)}$ from the 
components $({b^a}_{\alpha})$. Accordingly we can put ${\it E(r)}=0$ without 
loss of generality. The metric tensor 
$g_{\mu \nu}$ is then written as
\ba
g_{0 0} \A=\A  -(C^2-H^2),  \nonu
g_{0 \alpha} \A=\A \{ -CG +DH \}n_\alpha, \nonu
g_{\alpha \beta} \A=\A (D^2+F^2) \delta_{\alpha \beta}-(F^2+G^2)
n_\alpha n_\beta. 
\ea

According to (30), the total energy of an isolated system can be calculated
if the superpotential is known up to order $O(1/r^2)$. It is then enough
to know the parallel vector fields up to order $O(1/r)$. So let us  
restrict our attention to the weak field far from the source, and analyze
the field equation in vacuum up to order $O(1/r^3)$. This has been performed in G${\rm R,}^{21)}$ showing quite generally that the gravitational mass is equal
 to the total energy for any stationary isolated system. In the NGR, however, 
the analysis of the field equation in vacuum has not yet been done for the 
general
case of $(a_1+a_2) \neq 0$ even in the weak field approximation. This is due to
the fact that the gravitational field equation of the NGR is more complicated 
than the Einstein equation of GR.

Let us suppose that the leading term of the five unknown functions is given by
 some power of $1/r$, and that ${(b^k}_\mu)$ can be represented as \footnote{
The constant ${\it b}$ should not be confused with det$({b^k}_\mu)$, which we 
denote by $\sqrt{-g}$.}
\be
\left({b^k}_\mu \right)= 
\left(
\matrix{
(1+\displaystyle{b \over r^s}) & \displaystyle{j \over r^t} n^\alpha 
\vspace{3mm} \cr 
\displaystyle{h \over r^u} n^a & (1+\displaystyle{d \over r^v})
{\delta_\alpha}^a +\displaystyle{f \over r^w}\epsilon_{a \alpha \beta} n^\beta 
\cr}
\right).
\ee
Here the powers, ${\it s}$, ${\it t}$, ${\it u}$, ${\it v}$ and ${\it w}$, are
positive unknown constants at the beginning of the calculation, and their value will be determined by the field equation of NGR. 
Here the constant coefficients, ${\it b}$, ${\it j}$, ${\it h}$, ${\it d}$ and
 ${\it f}$, are also unknown, but they can be assumed to be nonvanishing without loss of generality since the powers of $r$ are left unknown.
 Use (42) in (41) gives the asymptotic behavior of the metric tensor which 
involves linear and quadratic terms of the unknown constants, ${\it b}$, 
${\it j}$, ${\it h}$, ${\it d}$ and ${\it f}$: The linear 
terms are dominant if the powers satisfy the following inequality 
\be
s<2u, \qquad v<2t, \qquad v<2w,
\ee
which we call the condition of the linear approximation.

Now apply the vacuum field equation (9) to (42), assuming the condition (43). 
Keeping only the leading 
terms, which are shown to be linear in the five unknown constants due to (43),
 we get the nonvanishing components of $I_{\mu \nu}$; 
\ba
I_{0 0} \A=\A {-2 \over r^2} \left \{
\displaystyle{s(s-1)(a_1+a_2)b \over r^s}-
\displaystyle {v(v-1)(a_1-2 a_2)d \over r^v} 
\right \},\\ 
I_{\alpha 0} \A=\A 2(u+1)(u-2)\displaystyle {(a_1+a_2)h \over r^{u+2}}n_\alpha, \ea
\ba
I_{\alpha \beta} \A=\A {n_\alpha n_\beta \over r^2} \left \{
\displaystyle{s(s+2)(a_1-2 a_2)b \over r^s}-
\displaystyle{v(v+2)(a_1+4 a_2)d \over r^v} \right \} \nonu
\A \A
-{\delta_{\alpha \beta} \over r^2} \left \{
\displaystyle{s^2 (a_1-2 a_2)b \over r^s}-
\displaystyle{v^2  (a_1+4 a_2)d \over r^v} \right \}\nonu
\A \A
+(w+1)(w-2)\displaystyle{(a_1-\displaystyle{4 \over 9}
 a_3)f \over r^{w+2}} \epsilon_{\alpha \beta \gamma} n^\gamma.
\ea
Here it is important to notice that $I_{\alpha 0}$ and $I_{[\alpha \beta]}$ 
are dominated by the linear term irrespectively of the condition of the linear 
approximation. Also using (42) in (36), we get the leading terms for the appropriate components of the superpotential;
\be
{{\cal U}_0}^{0 \alpha} =- {2n^\alpha \over \kappa r}
\left[{s(a_1+a_2)b \over r^s}- 
{v(a_1-2 a_2)d \over r^v} \right].
\ee

By virtue of (44) and (46) we can show that the equations,  $I_{0 0}=0$ and 
$I_{(\alpha \beta)}=0$, give the following results: 

1) if $s<v$ then $b=0$, because $(a_1-2a_2)\neq 0$, 

2) if $s>v$ then $d=0$, because $(a_1+4a_2)\neq 0$, and 


3) if $s=v\neq 1$ then  $b=d=0$, because 

\be
{\rm det} \left(
\matrix{
(a_1+a_2) & -(a_1-2a_2)  \vspace{3mm} \cr 
(a_1-2a_2) & -(a_1+4a_2)
\cr}
\right) \neq 0.
\ee
All the above three cases contradict our assumption that $b\neq 0$ and 
$d\neq 0$. Therefore, we find that the powers $s$ and $v$ should be given by
\be
s=v=1.
\ee
Then the equation $I_{0 0}=0$ is automatically satisfied and the remaining 
one, 
$I_{(\alpha \beta)}=0$, gives
\be
d=(2\epsilon-1)b.
\ee
We notice that the asymptotic behavior of the diagonal exact solution (21) 
satisfy (49) and (50) with $b=-m$. Also this result of (49) is physically 
very satisfactory in view of the superpotential (47), because we can
 then get a finite value of the energy. 

Next the $(\alpha 0)$-component of the field equation in
 vacuum, $I_{\alpha 0}=0$, implies that
\be 
(u-2)(a_1+a_2)=0.
\ee
When $(a_1+a_2)\neq0$, this gives
\be
u=2, 
\ee
compatibly with the condition of the linear approximation (43).
 Using (49) and (51) in (41), we
see that the component $g_{00}$ of the metric tensor behaves asymptotically
like
\be
g_{0 0}=-\left( 1+\displaystyle{2b \over r} \right),
\ee
which indicates that the constant $c$ is related to the gravitational mass $m$  of the isolated system by
\be
b=-m,
\ee
since as shown in \S 2 the world line of a spinless test particle is a 
geodesics of the metric. So (50) can now be written as 
\be
d=(1-2\epsilon)m.
\ee
In the special case of $(a_1+a_2)=0$, on the other hand, the power $u$ is not 
constrained by the field equation $I_{\alpha 0}=0$. 
However, the exact form of all the spherically symmetric solutions was found in this special case, and the energy of those solutions was 
${\rm calculated,}^{23)}$ as will be explained at the end of this section.
 
Finally from the skew part, $I_{[\alpha \beta]}=0$, we get
\be
(w-2) \left(a_1-{4 a_3 \over 9} \right)=0.
\ee
When  $(a_1-4 a_3/9) \neq 0$, it follows from the
 condition (56) that
\be
w=2,
\ee
satisfying the condition of the linear approximation (43). 
In the special case of \\ $( a_1-4 a_3/9)=0$, on the other hand, the power $w$ is not restricted by the field equation. We shall discuss this case separately 
in the next section.

Collecting the above arguments together, we see that when the parameters satisfy $( a_1+ a_2) ( a_1-4a_3/9) \neq 0$, the field equation 
$I_{\mu \nu}=0$ can be solved in the linear approximation to give the following asymptotic form of $({b^k}_\mu)$:
\ba
\left({{b^k}_\mu} \right) \A = \A 
\left(
\matrix{
(1-\displaystyle{m \over r}) & 
j \displaystyle{n_\alpha \over r^t}  \vspace{3mm} \cr 
h \displaystyle{n^a \over r^2} & 
(1+\displaystyle{m (1-2\epsilon) \over r}){\delta^a}_{\alpha}+
{f \over r^2} \epsilon_{\alpha \beta \gamma} n^\gamma
\cr}\right) \nonu
\A=\A \left( {{ b^k}_{\mu}{\rm (exact)}} \right)+
\left(
\matrix{
0 & j \displaystyle{n_\alpha \over r^t}  \vspace{3mm} \cr 
0 & 0 \cr
}\right)+O \left({1 \over r^2} \right),
\ea
where $\left( {{ b^k}_{\mu}{(\rm exact)}} \right)$ denotes the exact solution 
(21) in the asymptotic form. Substituting (58) into (36) gives
\be
{{\cal U}_0}^{0 \alpha} = {2m n^\alpha \over \kappa r^2}
\left[(a_1+a_2)+(1-2 \epsilon) 
(a_1-2 a_2) \right]= -{2m n^\alpha \over \kappa r^2},
\ee
where we have used (16) in the last equation. From (30) we then get
\be
E=m.
\ee
Thus, the gravitational mass is equal to the total energy for an isolated 
spherically symmetric system.

For completeness let us recapitulate the known results$^{\rm 23)}$ about the 
exact, spherically symmetric solutions and their total energy. 
Referring to the general expression  for 
the parallel vector fields given by (39), the solutions are divided 
into two classes: (1) the solution with $F(r)=0$ and (2) the solution with 
$F(r) \neq 0$. The general solution of the class (1) involves an arbitrary 
function of $r$, and therefore there exist solutions whose components 
$({b^a}_0)$  asymptotically behave like $1/r^u$ for any positive value of $u$.
 The calculated energy was shown to coincide with the gravitational mass only 
when the components $({b^a}_0)$ go to zero faster than $1/\sqrt{r}$. When 
$({b^a}_0) \sim h n^\alpha/\sqrt{r}$ for large $r$, the quadratic terms of $h$
 must be taken into account in the superpotential, and the energy does not 
coincide with the gravitational mass, being given by
\be
E=m+{h^2 \over 2}.
\ee
If $({b^a}_0)$ behave like $1/r^u$ with $0<u<1/2$, the calculated energy is 
infinite, implying that the solutions with such an asymptotic behavior is
 physically unacceptable. Next let us turn to the general solution of the 
class (2), which was shown to involve an arbitrary constant parameter, and 
to have the asymptotic behavior of (42) with $h=0$ and $w=2$: The calculated 
energy was shown to coincide with the gravitational mass.

\newsection{The energy in the special case of $(a_1-4 a_3/9)=0$}
Now let us turn to the special case of $(a_1 -4 a_3/9)=0$, in which we must go
beyond the linear approximation discussed in the preceding section since the 
field equation does not impose any restriction on the power $w$. We notice that
when the parameters, $a_1$, $a_2$ and $a_3$ satisfy the conditions, 
$(a_1+a_2)=0$ and $(a_1-4a_3/9)=0$, together with the Newtonian approximation
 condition (15), the gravitational Lagrangian (5) reduces to the 
Einstein-Hilbert Lagrangian of GR written in terms of the tetrad field, thus 
leading to
 inconsistency of the field equation (9): For example, the $I_{\mu \nu}$ is 
symmetric, while $T_{\mu \nu}$ of spinor fields is nonsymmetric. Accordingly
 we assume $(a_1+a_2) \neq 0$ in this special case.

Now consider the parallel vector fields $({b^k}_\mu)$ which asymptotically 
behave like (42) with $u=2$. Since the power $w$ cannot be fixed by the 
field equation, we shall discuss the three cases, $w>1/2$, $w=1/2$ 
and $w<1/2$, separately.\\

\noindent
(i) When $w>1/2$, it can be shown by the same argument as in the previous section that the powers $s$ and $v$ satisfy either $s=v=1$, or $s>1$ and $v>1$. If 
 $s=v=1$, the linear approximation studied in the preceding section is still 
valid, and the parallel vector fields behave asymptotically like
\be
\left({{b^k}_\mu} \right)= \left( {{ b^k}_{\mu}(exact)} \right)+
\left(
\matrix{
0 & j \displaystyle{n_\alpha \over r^t}  \vspace{3mm} \cr 
0 &  \displaystyle{f \over r^w} \epsilon_{a \alpha \beta}n^\beta  \cr
}\right),
\ee
and the energy coincides with the gravitational mass.

If it happens that $s>1$ and $v>1$, on the other hand, the superpotential dies 
out at infinity faster than $1/r^2$, and hence the calculated energy will be vanishing. The gravitational mass is also vanishing since $g_{0 0}$ tends to 1 
for large $r$ faster than $1/r$. Therefore, such a solution, if it exists, is
 devoid of physical meaning, although we cannot exclude its existence by the  
present approximate treatment.\\

\noindent
(ii) When $w=1/2$, the field equation $I_{\alpha \beta}=0$ implies that the 
powers $s$ and $v$ must satisfy one of the three 
possible alternatives: (a) $s=v=1$, (b) $s=1$ and $v>1$ or (c) $s>1$ and 
$v=1$. 

Let us start with the case (a) of $s=v=1$. The left-hand side of (9) is given 
by 
\ba
I_{0 0}\A =\A  I_{0 \alpha}= I_{\alpha 0}= 0, \nonu
I_{\alpha \beta} \A=\A -\left[ b(a_1-2a_2)
- d(a_1+4a_2)+2f^2(a_1+a_2)
 \right] \left( \displaystyle{{\delta_{\alpha \beta} -3n_\alpha n_\beta} \over
 r^3} \right)
\ea
up to order $O(1/r^3)$, where we have used 
$(a_1-4a_3/9)=0$, and the constant $c$ satisfies (54) since $s=1$ and $u=2$. 
 The vacuum field equation then provides the condition
\be
d={1 \over a_1+4 a_2}
\left[ b(a_1-2 a_2) +2f^2(a_1+a_2) \right]
 = (1-2\epsilon)m+2\epsilon f^2.
\ee
Similarly the superpotential is expressed by
\ba
{{\cal U}_0}^{0 \alpha} \A=\A -{2 n^\alpha \over \kappa r^2}
\left[ (a_1+a_2)b- (a_1-2 a_2)d+{1 \over 2}(a_1-2 a_2)f^2 \right]\nonu
\A=\A -{2m n^\alpha \over \kappa r^2} \left[ 1-\displaystyle{1-2\epsilon \over 
1-\epsilon}\displaystyle{f^2 \over 2m} \right],
\ea
where we have used (16) as well as (64).  With the help of (30) we then have
\be
E=m-{1-2\epsilon \over 2(1-\epsilon)}f^2 .
\ee

In the case (b) of $s=1$ and $v>1$, the terms proportional to the constant $d$ 
do not contribute to the leading term of 
$I_{\alpha \beta}$ and ${{\cal U}_0}^{0 \alpha}$. Accordingly, the 
calculated energy is written as
\be
E={m \over 4\epsilon(1-\epsilon)},
\ee
which goes to infinity as $\epsilon$ tends to zero (i.e., 
$(a_1+a_2)\rightarrow 0$ ). This means that when the parameter $\epsilon$ is 
varied, the solution of the case (b), if it exists, will be singular at 
$\epsilon=0$. 

\noindent
The last case (c) of $s>1$ and $v=1$ can be examined in the similar manner.
 Since $s>1$, the gravitational mass of the central body must be vanishing, and furthermore the energy takes a value
\be
E=-{1-2\epsilon \over 2(1-\epsilon)}f^2,
\ee
which is negative due to the experimental value of $\epsilon$ given by (17). 
Thus, the case (c) should be discarded.

Therefore, when the power $w$ is equal to $1/2$, the energy of the spherically
 symmetric body differs from the gravitational mass as measured from far 
distances.  It must be noticed, however, that we have not yet actually found a 
solution with such asymptotic behavior with $w=1/2$. The above result
 merely implies that the existence of such a solution is not 
excluded by the field equation which considers only up to order $O(1/r^3)$:
 Detailed analysis based on exact solutions is desirable.

\noindent
(iii) When $w<1/2$, quadratic terms of $f$ contribute to the superpotential with  the asymptotic behavior like $1/r^{2w}$, and hence the energy integral will 
diverge, giving an infinite value. Thus, the solution with $w<1/2$, if it 
exists, must be discarded.
\newsection{Conclusion and discussion}

We studied the problem whether or not the equality of the active gravitational mass and the inertial mass (i.e., the total energy) is ensured by the gravitational field equation in NGR, restricting ourselves to isolated spherically 
symmetric systems.
Based on the identity following from the general coordinate invariance, we
 defined the gravitational energy-momentum 
complex and derived the explicit expression for the superpotential 
 which allows us to calculate the total energy of isolated systems.
 We first applied it to the exact spherically symmetric  solution of diagonal 
matrix form, 
and found that the equality of the active gravitational mass and the total 
energy is satisfied.

Since we do not know whether the spherically symmetric solution is 
unique, we then discussed the asymptotic method to calculate the total energy.  Namely, we started from a quite general expression for the parallel vector 
fields with spherical symmetry, and solved the vacuum field equation at far 
distances from the source. The main results can be summarized as follows:\\
(a) In the generic case of $(a_1+a_2) (a_1-4a_3/9)\neq0$, the linear 
approximation can be applied to solve the field equation. All the components of the parallel vector fields other than $({b^{(0)}}_\alpha)$ are asymptotically 
determined by the field equation, and coincide with those of the exact 
solution up to order $O(1/r)$. The components $({b^{(0)}}_\alpha)$  do not 
contribute to the total energy, however, and the calculated value of the energy agrees with the gravitational mass.\\
(b) In the special case of $(a_1+a_2)=0$ the asymptotic behavior of the components $({b^a}_0)$ is not fixed at all by the field equation. We know, however, that for any asymptotic behavior of $({b^a}_0)$, there exist exact solutions with 
such asymptotic behavior.$^{23)}$ The total energy is finite only when 
$({b^a}_0)\sim 1/r^u$  with $u\geq1/2$, and furthermore if $u=1/2$, the 
quadratic terms of
  $({b^a}_0)$ contribute to the total energy, violating the equality of the 
total energy and the gravitational mass.\\
(c) In the special case of $(a_1-4a_3/9)=0$, the gravitational field equation 
does not impose any condition on the $\epsilon_{a \alpha \beta} n^\beta$-term 
of $({b^a}_\alpha)$. We showed that if this term tends to zero at infinity like  $1/\sqrt{r}$, the equality between the gravitational mass and the total energy is violated. It is yet to be studied, however, whether exact solutions with such asymptotic behavior really exist.

Thus, in the special cases of (b) and (c) above, there are anomalous solutions 
which violate the equality of the energy and the gravitational mass. The 
characteristic feature of these solutions is that specific components of the 
parallel vector fields tend to zero as $1/\sqrt{r}$ for large $r$. The physical meaning of these anomalous solutions is not yet clear, however. Is it reasonable to rule out such anomalous solutions by demanding that all the components of 
the parallel vector fields should tend to flat-space value faster than 
$1/\sqrt{r}$?

As we noticed, if the parameters satisfy both the conditions, $(a_1+a_2)=0$ and   $(a_1-4a_3/9)=0$, then the gravitational Lagrangian in the NGR coincides with
 the Einstein-Hilbert Lagrangian in GR expressed in terms of the tetrad field.
 The gravitational field equation of the NGR then becomes inconsistent: 
In fact, 
the L.H.S. of (9) will be symmetric but the R.H.S. is nonsymmetric. Therefore,
 we assumed in this paper that the combinations of the parameters, $(a_1+a_2)$ 
and  $(a_1-4a_3/9)$, do not vanish at the same time. The special case of 
$(a_1+a_2)=0$ 
has been studied in rather detail.$^{3),4),9),22),23)}$ Another case of 
$(a_1-4a_3/9)=0$ deserves more attention in this respect.

We assumed the spherical symmetry to obtain the solution at large distances. In GR, however, it is well known that such assumption of spherical symmetry can be removed. In fact, Einstein equation ensures that the metric tensor at far distances is the same as the Schwarzschild metric up to order $O(1/r)$ for any isolated stationary system. Is it possible in NGR to investigate the vacuum solution at far distances without assuming spherical symmetry?

We have ignored the parity-violating term, $v_\mu$$a^\mu$, throughout this 
paper. The possibility that this term plays an important role was
suggested by M\"uller-Hoissen and ${\rm Nitsch}^{26)}$ in the case of 
$(a_1+a_2)=0$. It seems interesting to take into 
account this parity-violating term also in the case of $(a_1-4 a_3/9)=0$.

We shall address ourselves to these problems in future work.

\newpage
\bigskip
\bigskip

\centerline{\Large{\bf Acknowledgements}}

The authors would like to thank Prof.\ K. Hayashi for a careful reading
of the manuscript. Also
one of the authors (G.N.) would like to thank Japanese Government for 
supporting him with a Monbusho Scholarship and also wishes to express his deep 
gratitude to all the members of Physics Department at Saitama University, 
especially to Prof.\ T. Saso, Prof. Y. Tanii and Dr.\ S. Yamaguchi. 
\bigskip


\newpage

\centerline{\Large{\bf References}}

\bigskip

\begin{enumerate}

\item[{1)}] C. M$\phi$ller, 
{\it Mat.\ Fys.\ Medd.\ Dan.\ Vid.\ Selsk.\ } {\bf 1} (1961), 10.\

\item[{2)}] C. Pellegrini and J. Plebanski,
{\it Mat.\ Fys.\ Skr.\ Dan.\ Vid.\ Selsk.\ } {\bf 2} (1963), 4.\

\item[{3)}] K. Hayashi and T. Nakano,
{\it Prog.\ Theor.\ Phys.\ } {\bf 38} (1967), 491.\

\item[{4)}] C. M$\phi$ller,
{\it Mat.\ Fys.\ Medd.\ Dan.\ Vid.\ Selsk.\ } {\bf 39} (1978), 13.\

\item[{5)}] H. Meyer,
{\it Gen.\ Rev.\ Grav.\ } {\bf 14} (1982), 531.\

\item[{6)}] F.W. Hehl, J. Nitsch, and P. Von der Heyde, 
in {\it General Relativity and Gravitation}, A. held, ed. 
(Plenum Press, New York, 1980), Vol.\ {\bf 1},  pp.\ 329-355.\

\item[{7)}] K. Hayashi and T. Shirafuji, 
{\it Prog.\ Theor.\ Phys.\ } {\bf 64} (1980), 866, 883, 1435, 2222; {\bf 65} 
(1980), 525.\

\item[{8)}] D. S\'aez,
{\it Phys.\ Rev.\ } {\bf D27} (1983), 2839.\

\item[{9)}] K. Hayashi and T. Shirafuji, 
{\it Phys.\ Rev.\ } {\bf D19} (1979), 3524; {\bf D24} (1981), 3312. \

\item[{10)}] A. Einstein, 
 {\it Sitzungsber.\ Preuss.\ Akad.\ Wiss.\ } (1928), 217. 

\item[{11)}] S. Miyamoto and T. Nakano,
{\it Prog.\ Theor.\ Phys.\ } {\bf 45} (1971), 295.\

\item[{12)}] T. Kawai and N. Toma, 
{\it Prog.\ Theor.\ Phys.\ } {\bf 83} (1990), 1.\

\item[{13)}] K. Hayashi and T. Shirafuji,  
{\it Prog.\ Theor.\ Phys.\ } {\bf 84} (1990), 36.\

\item[{14)}] C.M. Will, 
{"\it Theory and Experiment in Gravitational Physics" } (Cambridge U.P., 
Cambridge, 1993).\

\item[{15)}] K. Nordtvedt, 
{\it Phy.\ Rev.\ } {\bf 169} (1968), 1014, 1017; {\bf 170} (1968), 1186.\

\item[{16)}] J.O. Dickey et al., 
{\it Science } {\bf 265} (1994), 482.\

\item[{17)}] J.G. Williams, X.X. Newhall and J.O. Dickey, 
{\it Phy.\ Rev.\ } {\bf D53} (1996), 6730.\

\item[{18)}] K. Nordtvedt, 
{\it Icarus } {\bf 114} (1995), 51.\

\item[{19)}] A.S. Eddington, 
{\it "The Mathematical Theory of Relativity"}  (Cambridge U.P., Cambridge, 1957),  p.\ 105.\

\item[{20)}] H.P. Robertson, 
in {\it"Space Age Astronomy"}, A.J. Deutsch and W.E. Klemperer, 
 ed.\ (Academic Press, New York, 1962), p.\ 228.\

\item[{21)}] C.W. Misner, K.S. Thorne and J.A. Wheeler,
{\it Gravitation} (Freeman, San Francisco, 1973), pp.\ 448-457.\

\item[{22)}] F.I. Mikhail, M.I. Wanas, A. Hindawi and E.I. Lashin,
{\it Int.\ J.\ Theor.\ Phys.\ } {\bf 32} (1993), 1627.\

\item[{23)}] T. Shirafuji, G.G.L. Nashed and K. Hayashi, 
{\it Prog.\ Theor.\ Phys.\ } {\bf 95} (1996), 665.\

\item[{24)}] C. M$\phi$ller, 
{\it Ann.\  Phy.\ } {\bf 4}, (1958) 347.\

\item[{25)}] H.P. Robertson, 
{\it Ann.\ Math.\ (Princeton)} {\bf 33} (1932), 496.\

\item[{26)}] F. M\"uller-Hoissen and J. Nitsch, 
{\it Phys.\ Rev.\ } {\bf D28}, (1983) 718.\

\end{enumerate}


\end{document}